# An ontology-based method for semantic integration of Business Components

Hicham Elasri, Abderrahim Sekkaki
Departement of Mathematics and Computer Science
University Hassan II, Ain Chock, Faculty of Sciences
Casablanca, Morocco
hicham_elasri@yahoo.com, a.sekkaki@fsac.ac.ma

Larbi Kzaz
Higher Institute of Commerce and Business Administration
(ISCAE) Km 9.5, Route de Nouasseur P.O Box. 8114 -
Casablanca Oasis. Morocco
kzaz_larbi@yahoo.fr

*Abstract*—**Building new business information systems from reusable components is today an approach widely adopted and used. Using this approach in analysis and design phases presents a great interest and requires the use of a particular class of components called Business Components (BC). Business Components are today developed by several manufacturers and are available in many repositories. However, reusing and integrating them in a new Information System requires detection and resolution of semantic conflicts. Moreover, most of integration and semantic conflict resolution systems rely on ontology alignment methods based on domain ontology. This work is positioned at the intersection of two research areas: Integration of reusable Business Components and alignment of ontologies for semantic conflict resolution. Our contribution concerns both the proposal of a BC integration solution based on ontologies alignment and a method for enriching the domain ontology used as a support for alignment.**

*Keywords-component; Business Components, Semantic Integration, Ontology alignment, Enriching Ontologies.*

I. INTRODUCTION

Developing new Business Information Systems (IS) from reusable components is today an approach widely adopted and used [1], [23], [25]. Using this approach includes implementation phases as well as preliminary phases of analysis and design. However, components needed during design and analysis phases are not technical but conceptual. In fact, this class of Components implements business logic and knowledge of a domain. Components involved in analysis and design phases are commonly referred to as Business Components (BC).

Many research have focused the last decade to design new IS from reusable components [1], [23]. Two ways of research in the area of the reuse are intensively explored. The first one called "design for reuse" is to develop methods and tools to produce reusable components. The second "design by reuse", is to develop methods and tools to exploit reusable components [34]. We are concerned in this research by the second way. Literature outlines several questions when we address the topic of designing a new Information system by reusing available components. In fact, the reuse of components requires several operations such: research, selection, adaptation, composition [33] and integration. This last operation has been identified by [1]; the author also points the axis of semantic integration. In fact, Integrating into the same IS of several business components which emanate from various sources produces different conflicts both syntactic and semantic. We focus in this work on detecting and resolving semantic name conflicts encountered during the integration process of business components [24].

We assume that the design of an IS intended generally a business domain and that business components model fragments of this domain. Otherwise, semantic integration systems are mostly based on the alignment of ontologies; this issue has given rise to several works [3] and [24]. We relied on results of these works to support semantic integration process and have proposed integration architecture based on the alignment of ontologies using domain ontology and a method of measuring semantic similarity [24].

However, this solution is insufficient when there is no direct semantic relation between concepts of the domain ontology (and / or when the concepts do not appear in the domain ontology) used to support alignment. To overcome this insufficiency, we propose to exploit results of some recent works using rules to enrich ontologies. In fact, the application of some rules to concepts which are candidate for alignment allows detecting new semantic relations. These new semantic relations will be then injected in the domain ontology in order to enrich it.

Business Components to integrate will be used as a basis for the generation of new semantic relations. An enrichment domain ontology phase will be added in our architecture and an extension of the method for calculating the similarity will be proposed. We will validate our results using a prototype that we have developed and tested on domain ontology and some BC.

Our paper is organized as follow: First the problem of semantic integration of BC is presented. In section 3 domain ontology based alignment and enrichment-rules based techniques are described. In Section 4 our proposal of BC semantic integration method is given, completed in section 5 by a domain ontology enrichment process. In section 6 an example of application and a prototype are presented in order to illustrate our proposal. Finally, section 7 presents the conclusion and perspectives of our work.

## II. SEMANTIC INTEGRATION OF BUSINESS COMPONENTS.

Components based approach is considered since earliest 1990's as a new information system development paradigm [1]. This approach aims to reduce significantly costs and cycle-time of developing software. Components based approach consists in building new systems by reusing available components. Using this approach in the earliest phases of system development presents a real interest [33]. According to this approach, a business IS will be built from a set of BC which are generally heterogeneous. In fact, these BC generally emanate from various sources. For example, a company trading IS could be designed from multiple BC such as: {"Sales", "Product", "Customer» etc. ..}.

Integrating many components coming from different sources into the same IS can give rise to different types of semantic conflicts. Several researchers [19], [20] identified three types of semantic conflicts: confusion, measure and name conflicts. We are interested in the present work exclusively to name conflicts.

Several research works and implementations have shown the interest and the potential applications of ontologies in the areas of software engineering, IS development [15], and semantic integration [3]. We rely on the results of this works to ensure detection and resolution of semantic conflicts between BC.

Building a business IS usually implies a management domain: Trade, Finance, Human Resources etc. Ontologies describing these management domains are now available [37]. Moreover, components to integrate describe fragments of business knowledge in a language chosen by their designers. Several studies have focused on the transformation of BC described in modeling languages such as UML to ontologies.

We have proposed [24] an integration architecture that reduces the problem of semantic integration of BC to a problem of ontologies alignment.

We use domain ontologies for multiple reasons: Firstly, domain ontologies describe concepts related to a domain, this corresponds fully with our problem, since the design of an IS intended generally a business domain. Secondly, domain ontologies are reusable inside the same domain [21], this property is very interesting to consider in BC reusing, which is the central aim of design by reuse approach.

## III. ONTOLOGIES ALIGNMENT AND ENRICHMENT.

Ontologies are recently initiated tools for structuring knowledge and are defined as a collection of concepts and their interrelationships, which provide an abstract view of an application domain. According to Gruber, ontology is defined as an explicit formal specification of terms of a domain and relations among them [35] [36].

Aligning ontologies consists in establishing semantic relations among concepts of various ontologies which describe the same field of knowledge. Aligning ontologies represents a great interest in application domains that manipulate heterogeneous knowledge, such as semantic web, communication in Multi-Agent Systems, data Waterhouse, schemas/ ontologies integration [14], etc. Several works on the alignment of ontologies have emerged over recent years; most of them are based on an external resource that can be either a general ontology or domain ontology [3], [14].

The enrichment of ontologies consists to evolve their semantic content in order to cover new knowledge and increase their semantic consistency. More precisely, the enrichment consists in identifying new items: concepts, terms and relationships, and then placing them in an existing ontology. Enrichment as well as manual construction of ontology turns out to be a tiresome and expensive work [6]; that's why several studies have proposed automated and semi-automated methods of enriching and building ontologies. All those methods rely on external sources from which new semantic knowledge are identified, evaluated and placed within the ontology to enrich. The sources can be unstructured text such as dictionaries, knowledge bases, semi-structured or structured data such as conceptual schemas [26]. The enrichment process ontology can be divided into two steps: a learning step to search for new concepts and relations, and a placing step to set concepts and relations within the ontology. Several works in the literature have been proposed to cover one and / or other of these steps [7] and [8]. Most of existing approaches, generally based on statistical and linguistic tools, have focused on adding new concepts and / or semantic relations.

In this paper we propose to enrich the domain ontology used for support the alignment of components ontologies. The purpose is to improve the efficiency of the similarity measuring method which is based on domain ontology; this will be achieved by adding new semantic relations.

## IV. BUSINESS COMPONENT INTEGRATION PROCESS.

Business Components provide services and / or data which are expressed in most cases, in a terminology freely chosen by their designers. Semantic integration of BC consists to attribute meaning to data and services in order to ensure their exchange between heterogeneous BC and thus to allow their integration into the same IS. We propose in this section an extension of the solution that we have presented previously in [24]. Our solution allows:

- Detection and resolution of semantic name conflicts among components business to integrate into the new IS.

- Production a new BC obtained from the integration of original business components.

- Enrichment of the used domain ontology during the semantic integration process.

Our proposal relies on the results of several research projects including those on the components transformation from a component modeling language into an ontology modeling language, and those related to the alignment and enrichment of domain ontologies [2], [3], [4], [5] [6], [16] and [17]. A global description is provided in the following figure:

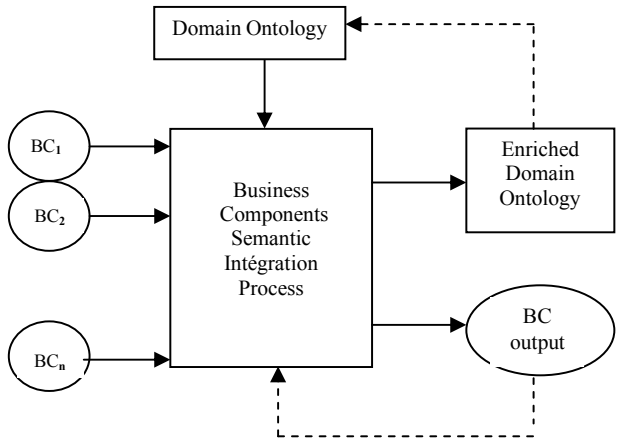

**Figure 1: Global view of Business Component integration Process**

The inputs of the integration process are:

- A set of Business Components selected by the designer in order to integrate them in the future Information system. We denote $BC_1….BC_n$, these BC.
- A domain ontology chosen by the designer according to the new IS domain. The domain ontology describes concepts and relations among concepts of the IS domain. The domain ontology will thereafter used to support the integration process.

Two outputs obtained at the end of the integration process:

- A new Business Component produced from the integration of the set of input BC.
- A domain ontology enriched by new semantic relations added by alignment and enrichment treatments.

The two outputs of the process may be subsequently reused in another iteration to integrate new BC:

- The new Business Component can be used by designers and architects in designing a new IS. This can be achieved by considering the new BC as a candidate to integrate with other components in another new IS.
- The enriched domain ontology can be used to update the original domain ontology and to support future integration process iterations. The use of the enriched domain ontology will increase the integration process efficiency.

The integration process involves the following steps:

1. Business Component transformation into Ontologies.
2. Ontologies alignment.
3. Ontology transformation into BC.

*A. Business Component transformation into ontologies.*

Several research studies have focused recently on the transformation of conceptual models described in a language such as UML into models using ontology description languages such as OWL. Thus [16] proposes a model driven (MDA) based methodology to generate ontologies from an annotated UML business model. Gasevic works [18] allow generating ontology from an UML model annotated by UML profile stereotypes of OWL provided by ODM (Ontology Definition Model). Transformations are performed by XSLT style sheet applied on XMI format models. [17] Presents an automated determining method of semantic relations among concepts of an ontology generated from UML conceptual models. We can also note the Eclipse project EODM6 (EMF Ontology Definition Meta model) which implements the standard ODM by using EMF (Eclipse Modeling Framework) technologies. EODM incorporates an inference layer and implements UML model transformation to RDFS and OWL models through Java APIs generated thanks to EMF. This transformation is shown below.

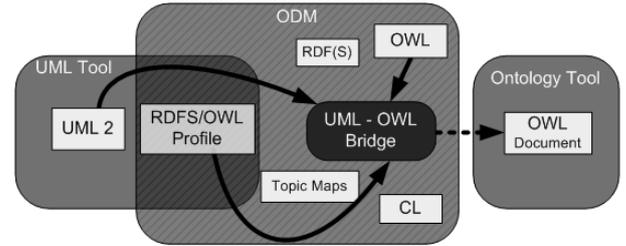

**Figure 2: UML model Transformation to OWL [38]**

Relying on the results of these studies, each BC candidate for integration is transformed into an ontology, thus bringing the problem of BC semantic integration to a problem of ontology alignment.

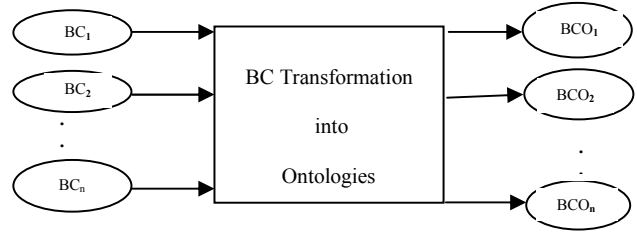

**Figure 3: Each $BC_i$ to integrate, is transformed into an ontology $BCO_i$**

*B. Ontologies alignment.*

- This step consists in aligning ontologies obtained from the transformation of BC. We use an alignment method based on domain ontology. This method is similar to ontologies alignment methods based on targeted complementary resources, also called background ontologies or support ontologies [2] [3] [18] and [19]. In our case, the ontology domain corresponds to that of the IS to design and from which BC to integrate are extracted. The domain ontology plays the role of targeted complementary resource and thus will be the support of ontologies alignment. This step of the process takes as input:

- A set of ontologies corresponding to each BC to integrate. These ontologies, denoted ($BCO_i$) in figure 4, are outputted from the last step.
- The domain Ontology chosen to support the alignment.

The outputs of this step are:

- An ontology, denoted BCOr in figure 4, resulting from the alignment of all BCOi ontologies submitted at input.
- An enriched domain ontology.

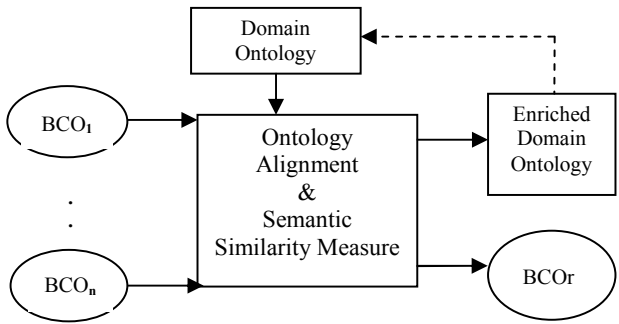

**Figure 4: Alignment of Business Component Ontologies**

In order to carry out alignment we propose a method of measurement of semantic similarity which will be given the responsibility to detect and to solve naming conflicts between concepts. The method of measurement of similarity semantics, noted σ thereafter, will be based on a method of measurement of syntactic similarity noted σ' like on a domain ontology noted Od.

That is to say Eci the set of concepts present in the ontology OBCi corresponding to the component BCi. That is to say EC the set of concepts present in all ontologies of components: EC. = Union (ECi) 1<=i<=n. Let be C1, C2 two concepts belonging to Ec. Let be *Term(Ci)* a function that returns the term used to describe the concept Ci

**Syntactic similarity measuring**

σ' is defined as follows:

σ': Ec ×Ec → {0, 1}

**begin**

  **if** C1 and C2 are atomic concepts then

     **if** Term(C1) =Term(C2) **then**  σ' (C1, C2) = 1

     **else**         σ' (C1, C2) = 0

     **endif**

   **else**

      % C1 and C2 are composites. C1 and C2 are then written C1 = (C11.., C1i,....., C1n) et C2 = (C21 ...., C2j,...., C2n) %

      σ' (C1, C2) =1/n (Σi j σ' (C1i, C2j)) 1 <= i, j <=n

   **endif**

**end**

The method σ' proposed, thus takes value 1 when the concepts are syntactically identical and 0 in the contrary case.

**Semantic similarity measuring.**

The method of measurement of the semantic similarity between concepts, is based on the domain ontology and uses the method of measurement of the syntactic similarity σ', defined here before. Are C1 and C2 two concepts of EC, Od the domain ontology, Rod the set of semantic relations available in Od, and R (C1, C2) the subset of the existing relations between the concepts C1 and C2 within Od. Rod ⊃ R (C1, C2.) σ the method of calculating the semantic similarity is defined as follows:

σ : $Ec \times Ec \rightarrow \{0, 1\}$,

**Inputs :**
- The two concepts C1 and C2 to compare semantically.
- The domain ontology $O_D$

**Outputs :** **1** if C1 and C2 are synonymous similar, **0** otherwise.

```
begin
 if (C1 and C2 ∈ O_D) then
    if (R (C1, C2) = ∅ ) then
       Start the ontology enrichment
       process.
       if a new relation is detected after
       the enrichment
       then     Update Rod and R (C1,C2)
                Recall σ (C1, C2)
       else % semantic similarity Measure
            coincides with the syntactic
            similarity measure %
            σ (C1, C2)= σ' (C1, C2)
    endif
    else
        if R (C1, C2) ⊃ a synonymous
                       relation
        then    σ (C1, C2) = 1
        else
            if R (C1, C2) ⊃ an
                 homonymous relation
            then σ (C1, C2) = 0
            else
                 σ (C1, C2)= σ' (C1, C2)
            endif
         endif
      endif
  else
   % C1 or C2 do not belong to OD the
   semantic    similarity    Measure
   coincides    with    the   syntactic
   similarity measure %
         σ (C1, C2)= σ' (C1, C2)
   endif
end.
```

*C. Transformation of result ontology of alignment into a new business component*

This is the last step of the integration process, it simply consists in converting the ontology BCOr, resulting from the alignment. This step produces a new Business Component, noted BCr in figure 5. The new BC can be used by designers and architects in designing a new IS. This can be achieved by considering the new BC as a candidate to integrate with other components in another new IS.

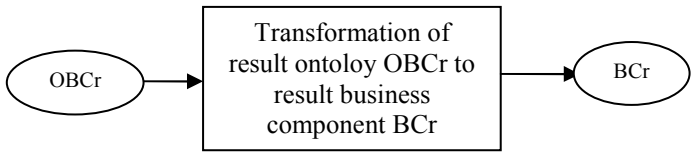

**Figure 5: Transformation of result ontology to result business component**

According to work [29], [31] and [32] it proves to be possible to transform ontologies towards UML. Result ontology OBCr is consequently transformable to a business component represented in UML. With this stage of the process of integration two ways arise: Automatic transformation of result ontology (OBCr) into a business component BCr result which will be integrated thereafter in the future IS. The second way consists first to use result ontology to detect and solve the semantic conflicts and then to manually transform OBCr into one business component BCr result which will be integrated in the future IS. This last way joined the proposals of [30] for the use of ontologies like means of assistance to the design of IS.

## V. ENRICHMENT PROCESS OF THE DOMAIN ONTOLOGY.

We propose to use an enrichment process of ontology, when this last contains no semantic relation between concepts to align. The enrichment process was in fact intended to discover possible semantic relations between these concepts. In order to implement this treatment and to prove its feasibility, we retained, as example, two rules among the various semantic relation rules:

- **R1: Two concepts are similar if their nearby equivalent concepts are similar.**

Indeed, according to [27] "Two concepts are similar if their sub-concepts are the same", so two concepts are similar if their "child" sub-concepts are the same. This rule was confirmed in [28].

- **R2: Two concepts are similar if their "child" sub-concepts are similar.**

This rule applies to composite concepts. Composite concepts represent the parent concepts and the sub-concepts, linked by part-of semantic relation type, are the child concepts.

Let be C1 and C2 the concepts to be aligned and OBCi the local ontology they belong to; we distinguish three cases:

**Case n° 1:** C1 and C2 admit a semantic relation within OBCi. This relation is then injected into the domain ontology OD.

**Case n° 2:** C1 and C2 do not admit a semantic relation in OBCi whereas there exists in OBCi two concepts C'1 and C'2 well as two semantic relations of equivalence; the first between C1 and C'1 and the second between C2 and C'2. According to R1 we can deduce a new semantic relation between C1 and C2 that one injects into the domain ontology OD.

**Case n° 3:** C1 and C2 are composite concepts which do not admit a semantic relation in OBCi, whereas there exist semantic relations between their "child" respective sub-concepts. Let be {C11, C12… C1n} the set "child" sub-concepts of C1 and {C21, C22… C2n} the set of "child" sub-concepts of C2 such that C1i and C2i admit a semantic relation within OBCi. According to R2 we can deduce a new semantic relation between C1 and C2 that one injects into the domain ontology OD.

## VI. ILLUSTRATION AND VALIDATION PROTOTYPE.

In order to validate our proposal, we give an example followed by a prototype which we have developed.

### D. EXAMPLE.

The example is based on a fragment of ontology (figure 6) and two components (figures 7 and 9) all relating to the field of "medical visits management". The fragment of ontology represents the domain ontology which will be used to support the semantic integration process. The business components noted BC1 and BC2, described in UML, represent the components candidates to semantic integration.

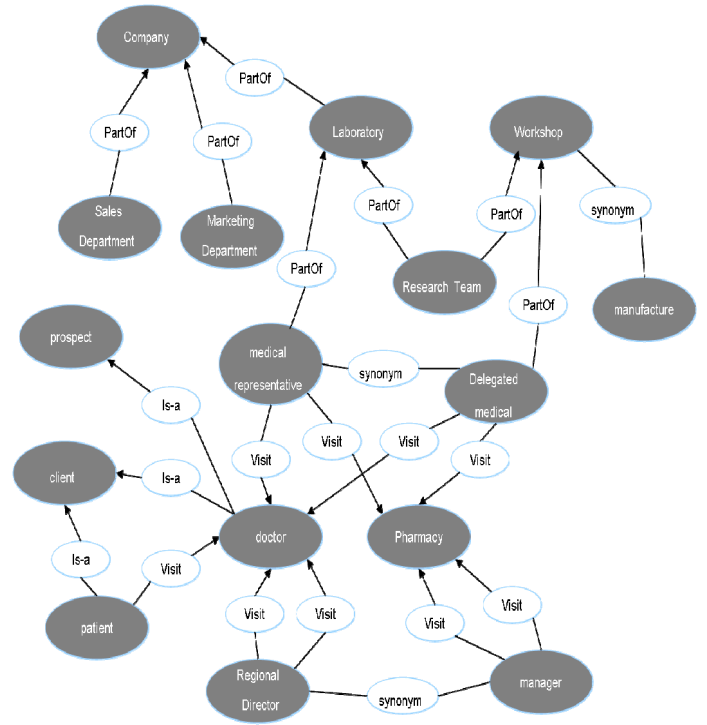

**Figure 6: Fragment of the "medical visits management" domain ontology.**

**Step n°1: Transformation of BC1 and BC2 into ontologies.**

We apply the transformations recommended in [17] and [18] for the transformation of BC1 (resp. BC2) towards OBC1 (resp. towards OBC2). The following table gives examples of correspondences between UML concepts and OWL concept.

| Business component (UML) | Ontology (OWL) |
|---|---|
| Class | Class |
| Association (ex : Agregation) | Property(ex : someValuesFrom) |
| Attribute | |
| DataType | DataType |

**Table 1: correspondence between UML and OWL**

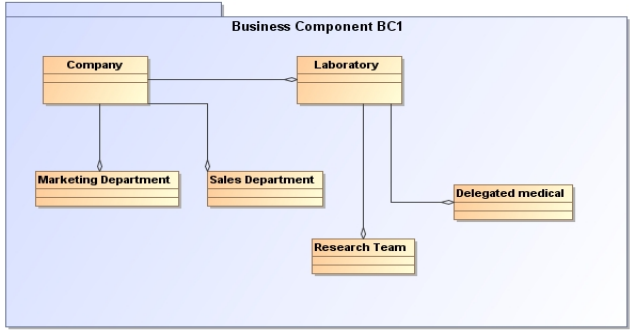

**Figure 7: First Business Component BC1 to integrate.**

The transformation of BC1 into ontology generates the ontology OBC1 hereafter:

```
Ontology(OBC1
 (Class Marketing Department partial
restriction(partOfsomeValuesFrom(Company))
(Class Sales Department partial
restriction(partOfsomeValuesFrom(Company))
(Class Laboratory partial
restriction(partOfsomeValuesFrom(Company))
(Class Delegated medical partial
restriction(partOfsomeValuesFrom(Laboratory))
(Class Research Team partial
restriction(partOfsomeValuesFrom(Laboratory)))
```

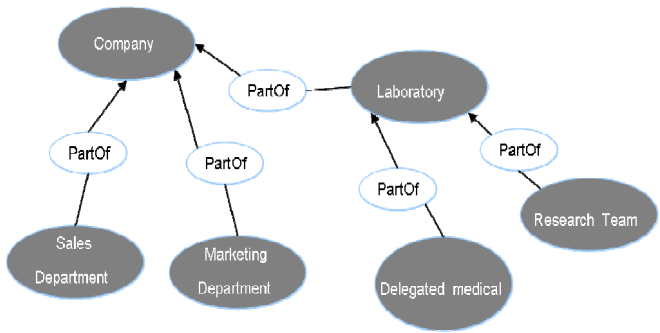

**Figure 8: Ontology OBC1 generated from component BC1**

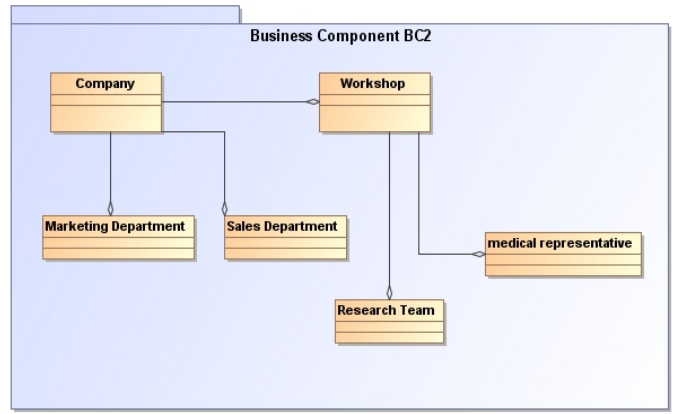

**Figure 9: Second Business Component BC2 to integrate.**

The transformation of BC2 into ontology generates the ontology OBC2 hereafter:

```
Ontology(OBC2
 (Class Marketing Department partial
restriction(partOfsomeValuesFrom(Company))
(Class Sales Department partial
restriction(partOfsomeValuesFrom(Company))
(Class Workshop partial
restriction(partOfsomeValuesFrom(Company))
(Class medical representative partial
restriction(partOfsomeValuesFrom(Workshop))
(Class Research Team partial
restriction(partOfsomeValuesFrom(Workshop)))
```

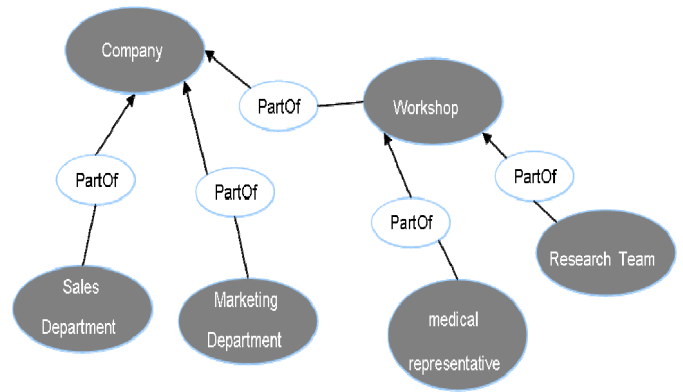

**Figure 10: ontology OBC2 generated from component BC2**

**Step n° 2: semantic integration and obtaining OBCr with highlighting the enrichment process**.

Ontology OBC1 generated from component BC1 comprises a concept called "Laboratory". Ontology OBC2 resulting from the component BC2 comprises a concept called "Workshop". The two concepts belong to the domain ontology.

(C1 and C2∈ OD) without admitting semantic relation between them (R (C1, C2) =∅). The alignment of the two concepts requires consequently "applying the enrichment process to the domain ontology". The two concepts having child sub-concepts "Medical Representative" and "Research team" are similar, according to R2 rule one can deduce that

"Laboratory" and "Workshop" are synonymous. A new relation "synonymy" is detected then added to the domain ontology. The calculation of σ ("workshop", "laboratory") then gives value 1. Thus the concepts "Laboratory" and "Workshop" thus will be linked by the synonymy type semantic relation. This relation is then added in OBCr ontology. Figure bellow presents the result of this processing.

```
Ontology(OBCr
(Class Marketing Department partial
restriction(partOfsomeValuesFrom(Company))
(Class Sales Department partial
restriction(partOfsomeValuesFrom(Company))
(Class Laboratory partial
restriction(partOfsomeValuesFrom(Company))
(Class Delegated medical partial
restriction(partOfsomeValuesFrom(Laboratory))
(Class Research Team partial
restriction(partOfsomeValuesFrom(Laboratory)))
(Class Workshop partial
restriction(partOfsomeValuesFrom(Company))
(Class medical representative partial
restriction(partOfsomeValuesFrom(Workshop))
(Class Research Team partial
restriction(partOfsomeValuesFrom(Workshop)))
(EquivalentClass(Laboratory, Workshop)
(EquivalentClass(Delegated medical, medical
representative ))
```

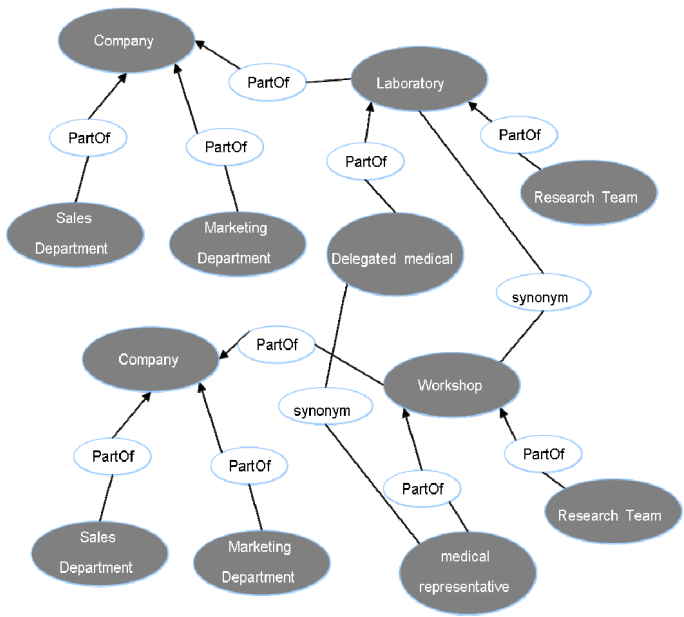

**Figure 11: result ontology OBCr**

**Step n° 3: Obtaining the integration process result BCr.**

At this step, designers can, as appropriate:

- Rely on OBCr ontology to note that BC1 and BC2 are synonymous; and to then choose BC1 or BC2 to use it in their new IS.
- Automatically transform OBCr ontology into a business component BCr. Figure bellow describes the resulting component BCr.

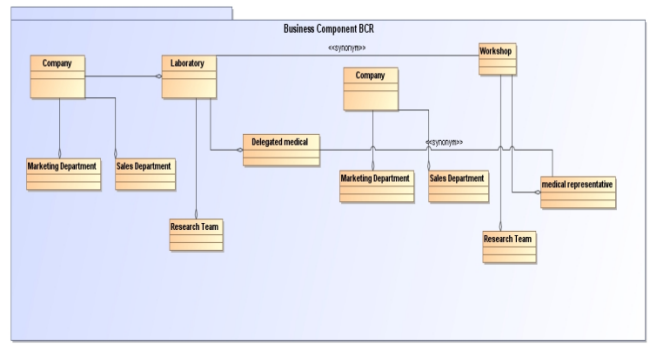

**Figure 12: The Business Component BCr resulting from integration**

*E. PROTOTYPE*

In order to validate and to evaluate our method of semantic integration, we have developed, using Growl project [9], a prototype baptized IntegrateBusinessOnto. This prototype can be seen like an extension of the Growl project. The Growl project is open source; it's developed in java and allows visualization and edition of ontologies described in OWL language. It allows users not specialists to easily manage ontologies. Our choice of Growl can be justified for the following reasons:

1- Growl allows the management of ontologies.
2- It is Open source.
3- It includes the popular API Jena for the managing ontologies.
4- It allows the visualization of ontologies
5- It is developed in JAVA, which enables interaction and easy integration with other java API's developed for ontology management.

The prototype IntégrateBusinessOnto takes first in entry the domain ontology and the set of business components to integrate; transforms them into ontologies described in OWL. Then, it applies the various treatments associated with each step of our proposed method, in particular similarity measurement and domain ontology enrichment. Finally IntégrateBusinessOnto outputs the resulting ontology described in OWL and its graphical description (figure 11).

VII. CONCLUSION AND PROSPECTS.

Our research lies within the scope of information systems engineering by re-use. We were interested more precisely in the resolution of semantic conflicts of naming type encountered during the re-use of business components in the analysis and design phases of new information systems. Our proposal is an application of domain ontologies to design IS by re-using conceptual business components; it consists of a three-step process. The first and last step concern the transformation of conceptual representations of business components into ontological representations and reciprocally. The second step, which constitutes the fundamental part of our work, consists of a method of calculating semantic similarity; it is based on the results of recent works on the ontologies alignment and the enrichment methods based on domain ontologies. An example of application has illustrated our proposal. We have also developed a prototype in order to validate the solution. The results obtained by the prototype implementation on some examples are encouraging. We think firstly to continue this work by a formal validation of the solution, and then by the research of the possibilities of extending it to solve other types of semantic conflicts, in particular measurement and confusion conflicts.